\documentclass{webofc}
\usepackage[varg]{txfonts}   
\newcommand{\be}{\begin{equation}}
\newcommand{\ee}{\end{equation}}

\newcommand{\lt}{\left}
\newcommand{\rt}{\right}

\newcommand{\non}{\nonumber \\}

\newcommand{\LMS}{\Lambda_{\overline{\rm MS}}}

\begin{document}
\title{Determination of HQET nonperturbative matrix elements with renormalon subtraction using Fourier transform}
%
%

\author{\firstname{Hiromasa} \lastname{Takaura}\inst{1}\thanks{\email{htakaura@post.kek.jp}}         
}

\institute{KEK Theory Center, Tsukuba 305-0801, Japan          }

\abstract{%
As higher order perturbative series are available, it is becoming necessary
to include nonperturbative effects in QCD calculations using the OPE.
In order to systematically determine nonperturbative effects and to incorporate them
into theoretical calculations, the renormalon problem should be resolved.
We use a renormalon subtraction method utilizing Fourier transform
to determine nonperturbative matrix elements of HQET, $\bar{\Lambda}$ and $\mu_{\pi}^2$.
This is the first determination performed with subtraction of the $u=1$ renormalon. 
}
\maketitle
\section{Introduction}
\label{intro}
Today high-order perturbative series at $\mathcal{O}(\alpha_s^3)$ and $\mathcal{O}(\alpha_s^4)$ 
are becoming available for many QCD observables such as the Adler function \cite{Baikov:2012zn}, 
the static QCD energy \cite{Brambilla:1999xf,Anzai:2009tm,Smirnov:2009fh}, 
the relation between the pole and the ${\overline{\rm MS}}$ masses \cite{Marquard:2016dcn},
and the semileptonic decay width of the B meson \cite{Fael:2020tow}.
These high-order results are improving precision of QCD calculations.
On the other hand, there are nonperturbative effects in QCD. 
These subleading effects are becoming relevant because of high precision calculation 
of  leading QCD corrections, i.e., perturbative calculations.
They can be formally included by using the framework of the OPE.
For instance, the semileptonic decay width of the B meson is given by
\be
\Gamma \approx \frac{G_F^2 |V_{cb}|^2}{192 \pi^3} m_b^5 \lt[C_{\bar{Q}Q}
+C_{\rm kin} \frac{\mu_{\pi}^2}{m_b^2}
+C_{\rm cm}  \frac{\mu_G^2}{m_b^2}+\cdots  \rt]  ,
\ee
where $\mu_{\pi}^2$ and $\mu_G^2$ [of $\mathcal{O}(\LMS^2)$] give nonperturbative corrections
to the perturbative contribution $C_{\bar{Q}Q}$.
In the 1S mass scheme, the perturbative contribution and
a nonperturbative correction are given by \cite{Hayashi:2022hjk}
\begin{align}
&\frac{m_b^5}{m_{b,{\rm 1S}}^5}C_{\bar{Q}Q}=0.5863|_{\rm LO}-0.0797|_{\rm NLO}-0.0294|_{\rm NNLO}-0.0073|_{\rm NNNLO}+\cdots  , \non
&C_{\rm cm} \frac{\mu_G^2}{(m_b^{\rm 1S})^2} \approx -0.0151 .
\end{align}
It turns out that the magnitude of the nonperturbative effect is larger than that of the highest order (NNNLO) perturbative contribution.
This observable is thus a clear example where nonperturbative effects 
are required to be included.

However, generally speaking, inclusion of nonperturbative effects is a non-trivial task.
This is because of the renormalon problem.
Let us consider the Adler function for a simple illustration.
The OPE is given by
\be
D(Q^2)=C_1(Q^2)+C_{FF}(Q^2) \frac{\langle FF \rangle}{Q^4}+\cdots .
\ee
The perturbative contribution $C_1$ is given by a divergent parturbative series
and the ambiguity related to the divergence is obtained as an imaginary part $\delta C_1 \approx \pm i \LMS^4/Q^4$
when the Borel resummation is applied.
This is called the $u=2$ renormalon uncertainty and is considered to cancel
against the uncertainty of the gluon condensate $\delta (\langle FF \rangle)/Q^4$.
This argument means that both quantities are ill-defined
unless some regularization is applied.
It particularly means that a fixed order perturbative result
is not sufficient to determine (or define) the nonperturbative effect.

In order to determine the nonperturbative effect (and incorporate it into QCD calculations),
we need to decompose the perturbative contribution into its real (renormalon-free) part
and imaginary part:
\be
C_1(Q^2)=C_{1, {\rm RF}}(Q^2) +\delta C_1(Q^2) .
\ee
Once this can be done, by assuming the cancellation of the perturbative and nonperturbative uncertainties,
one can rewrite the OPE such that each term is given by a renormalon-free (real-valued) quantity:
\be
D(Q^2)=C_{1, {\rm RF}}(Q^2)+C_{FF}(Q^2) \frac{\langle FF \rangle_ {\rm RF}}{Q^4}+\cdots .
\ee
This equation allows one to determine $\langle FF \rangle_ {\rm RF}$ (by using, for instance, 
the experimental result of the Adler function). Note that the gluon condensate appears not only for the Adler function
but also other observables such as the gradient-flow action density.
Therefore once $\langle FF \rangle_ {\rm RF}$ is determined by using a observable,
that value can be an input to calculate other observables and allows one to realize nonperturbative precision.

To this end, one needs to perform the decomposition of $C_1$ practically 
based on finite order perturbative results.
There are some proposed methods to do this \cite{Lee:2002sn,Ayala:2019uaw,Takaura:2020byt}.
In this paper, I explain the method of ref.~\cite{Hayashi:2021vdq} and consider its application
to a determination of HQET nonperturbative matrix elements.

This paper is organized as follows. In sec.~\ref{sec:2},
we explain the OPE of the B and the D meson masses
and renormalon cancellation in this observable.
To determine the nonperturbative matrix elements appearing in the OPE,
we need to decompose the pole mass, which can be calculated perturbatively
through the relation between the pole and the ${\overline{\rm MS}}$ masses,
into its renormalon-free part and renormalon uncertainties.
The calculation method to perform this decomposition is explained in sec.~\ref{sec:3}.
We give our results for the nonperturbative matrix elements in sec.~\ref{sec:4}.
Sec.~\ref{sec:5} is devoted to the conclusion.

\section{B and D meson masses in HQET}
\label{sec:2}

We consider the spin-averaged B and D meson masses:
\be
\overline{M}_B \equiv \frac{M_B+3 M_{B^*}}{4}, \quad{}
\overline{M}_D \equiv \frac{M_D+3 M_{D^*}}{4} .
\ee
The OPE formulas are given by
\begin{align}
\overline{M}_B&=m_b+\bar{\Lambda}+\frac{\mu_{\pi}^2}{2 m_b} +\cdots, \non
\overline{M}_D&=m_c+\bar{\Lambda}+\frac{\mu_{\pi}^2}{2 m_c} +\cdots . \label{HQETOPE}
\end{align}
Here $m_b$ and $m_c$ are pole masses of the charm and the bottom quarks, respectively.
$\bar{\Lambda}$ is the $\mathcal{O}(\LMS)$ nonperturbative matrix element,
and $\mu_{\pi}^2 \approx \langle B(p) | \bar{b}_v D_{\perp}^2 b_v | B(p) \rangle/(2 m_B)
\approx \langle D(p) | \bar{b}_v D_{\perp}^2 b_v | D(p) \rangle/(2 m_D)$ is the 
$\mathcal{O}(\LMS^2)$ nonperturbative matrix element.
The heavy quark symmetry tells us that $\bar{\Lambda}$ has 
an identical value for $B$ and $D$ mesons.
This applies also to $\mu_{\pi}^2$ (up to the order this study concerns).

The pole masses have the $u=1/2$ and the $u=1$ renormalon uncertainties,
and can be formally decomposed as
\be
m_{b(c)}=m_{b(c), {\rm RF}} \pm i \mathcal{O}(\LMS) \pm i \mathcal{O}(\LMS^2/m_{b(c), {\rm RF}}) +\cdots . \label{polemass}
\ee
Once this decomposition can be done, we have the renormalon-free OPE as
\begin{align}
\overline{M}_B&=m_{b, {\rm RF}} +\bar{\Lambda}_{\rm RF}+\frac{\mu^2_{\pi, {\rm RF}}}{2 m_{b, \rm RF}} +\cdots, \non
\overline{M}_D&=m_{c, \rm RF}+\bar{\Lambda}_{\rm RF}+\frac{\mu^2_{\pi, {\rm RF}}}{2 m_{c, \rm RF}} +\cdots . \label{RFOPE}
\end{align}

We give eq.~\eqref{polemass} or more explicitly $m_{b, {\rm RF}}$ and $m_{c, {\rm RF}}$.
To this end, we use the relation between the pole and the ${\overline{\rm MS}}$ masses
and subtract the $u=1/2$ and the $u=1$ renormalons in this relation.
We subtract the $u=1$ renormalon for the first time.
Then using the input ${\overline{\rm MS}}$ masses we can give $m_{b, {\rm RF}}$ and $m_{c, {\rm RF}}$.
Finally using the experimental mass values of the B and D mesons,
we determine $\bar{\Lambda}_{\rm RF}$, $\mu^2_{\pi, {\rm RF}}$ based on eq.~\eqref{RFOPE}.

\section{Renormalon subtraction using Fourier transform from the mass relation}
\label{sec:3}

The relation between the pole and the ${\overline{\rm MS}}$ masses is given by
\be
r(\overline{m}) \equiv \frac{m}{\overline{m}}
=1+c_0 \alpha_s(\mu^2)+(c_1+c_0 b_0 \log(\mu^2/\overline{m}^2)) \alpha^2_s(\mu^2)+\cdots , \label{usualseries}
\ee
where $c_0$, $c_1$, ... are log independent constants and 
$\mu^2 d \alpha_s/d \mu^2=\beta(\alpha_s)=-b_0 \alpha_s^2-b_1 \alpha_s^3-\cdots$.
The perturbative series has the $u=1/2$ and the $u=1$ renormalons.

We convert $r(\overline{m})$, a function of $\overline{m}$, to a function of 
the dual space variable $\tau$ by Fourier transform:
\be
\tilde{r}(\tau) \equiv
\frac{\pi}{\tau} \int_0^{\infty} \frac{d \overline{m}}{\overline{m}}
\sin (\tau/\sqrt{\overline{m}}) r(\overline{m}) . \label{dual}
\ee
The mass dimension of $\tau$ is $1/2$.
One can show that the $u=1/2$ and the $u=1$ renormalons originally encoded in
$r(m)$ are suppressed in $\tilde{r}(\tau)$.
Let us assume that renormalon uncertainties are given by form $\delta r(\overline{m})=\lt(\LMS^2/\overline{m}^2 \rt)^u$,
where $u=1/2$ ($u=1$) corresponds to the $u=1/2$ ($u=1$) renormalon.
The uncertainties of $\tilde{r}(\tau)$ induced by $\delta r(\overline{m})=\lt(\LMS^2/\overline{m}^2 \rt)^u$
are given by
\begin{align}
\delta \tilde{r}(\tau)
&=\frac{\pi}{\tau} \int_0^{\infty} \frac{d \overline{m}}{\overline{m}}
\sin (\tau/\sqrt{\overline{m}}) \delta r(\overline{m})   \non
&=\frac{\pi}{\tau} \int_0^{\infty} \frac{d \overline{m}}{\overline{m}}
\sin (\tau/\sqrt{\overline{m}}) \lt(\LMS^2/\overline{m}^2 \rt)^u   \non
&=2 \lt(\frac{\LMS}{\tau^2} \rt)^{2u} \sin(2 \pi u) \Gamma(4u) .
\end{align}
This is zero for $u=1/2$ and $u=1$ due to the sine factor.
Therefore $\tilde{r}(\tau)$ is free from the $u=1/2$ and the $u=1$ renormalons.

The inverse transform to obtain $r(\overline{m})$ from $\tilde{r}(\tau)$
is given by
\be
r(\overline{m})
=-\frac{1}{2\pi^2 \sqrt{\overline{m}}}
\int_0^{\infty} d\tau 
\sin(\tau/\sqrt{\overline{m}}) \tilde{r} (\tau)  .\label{inverse}
\ee
The left-hand side has the renormalons while the integrand on the right-hand side
is free of them.
Then the renormalons stem from the integration over $\tau$.
Particularly they stem from the integration of multiple logarithms
$\log(\mu^2/\tau^4)$, where the perturbative coefficient of $\mathcal{O}(\alpha_s^{n+1})$
of $\tilde{r}(\tau)$, which is RG invariant, is given by the $n$th order polynomial of $\log(\mu^2/\tau^4)$.
\be
\tilde{r}(\tau)=\frac{\pi^2}{\tau}
[1+\tilde{c}_0 \alpha_s(\mu^2)+(\tilde{c}_1+\tilde{c}_0 b_0 \log(\mu^2/\tau^4)) \alpha_s^2(\mu^2)+\cdots] . \label{dualseries}
\ee

We can avoid regeneration of the renormalons as follows.
We apply RG improvement to $\tilde{r}(\tau)$:
\be
\tilde{r}(\tau)=\frac{\pi^2}{\tau}[1+\tilde{c}_0 \alpha_s(\tau^4)+\tilde{c}_1 \alpha^2_s(\tau^4)+\cdots] . \label{RGdualseries}
\ee
Then in eq.~\eqref{inverse} the Landau pole singularity of $\alpha_s(\tau^4)$
is the only obstacle to performing the integral.
Therefore we can give a renormalon-free (real-valued) result using the principal value (PV) regularization:
\be
r(\overline{m})_{\rm RF}
=-\frac{1}{2\pi^2 \sqrt{\overline{m}}}
\int_{\rm PV} d\tau 
\sin(\tau/\sqrt{\overline{m}}) \tilde{r} (\tau) . \label{PVint}
\ee
Here the PV regularization means taking the average over the integrations along the paths $\tau=0 \pm i0 \to \infty \pm i0$.
The equivalence of $r(\overline{m})_{\rm RF}$ defined in the above way to the PV result of the Borel integral
is discussed in \cite{Hayashi:2022hjk}.
(Note that the PV result of the Borel integral means the PV regularization of the integral over the Borel variable $u$,
and thus this equivalence is not trivial.)

To summarize, our calculation procedure is as follows. 
We start with the usual fixed-order perturbative series as written in eq.~\eqref{usualseries}.
Then we calculate the dual-space perturbative series according to eq.~\eqref{dual}
to obtain the series of the form \eqref{dualseries}.
We can obtain the series exactly up to the order where the original series is calculated. 
Then we perform the PV integration over $\tau$ [eq.~\eqref{PVint}]
with the RG improved series eq.~\eqref{RGdualseries}.

We make a remark on finite mass corrections to the original-space series.
We regard $b$ and $c$ heavy quarks. 
This assures the flavor universality of the matrix elements.
In accordance with this, we should use the perturbative series for the mass relation $m_b/\overline{m}_b$ 
which contains finite charm mass corrections.
We also include non-decoupling effects of the bottom quark on $m_c/\overline{m}_c$.
In this case the perturbative series are given by \cite{Fael:2020bgs,Hayashi:2021vdq}
\begin{align}
m_b/\overline{m}_b
&=1+0.424413 \alpha_s^{(3)}(\overline{m}_b^2)+1.03744 (\alpha_s^{(3)}(\overline{m}_b^2))^2
+3.74358 (\alpha_s^{(3)}(\overline{m}_b^2))^3+\cdots, \non
m_c/\overline{m}_c
&=1+0.424413 \alpha_s^{(3)}(\overline{m}_c^2)+1.04375 (\alpha_s^{(3)}(\overline{m}_c^2))^2
+3.75736 (\alpha_s^{(3)}(\overline{m}_c^2))^3+\cdots , \label{masscorr}
\end{align}
where $\alpha_s^{(3)}$ represents the three-flavor coupling.
They exhibit a very similar behavior, and this is consistent with 
an implication of the OPE \eqref{HQETOPE} that
the $u=1/2$ renormalon uncertainties of $m_b$ and $m_c$
should be the same so that they can be absorbed by the single parameter $\bar{\Lambda}$.
In contrast, the naive perturbative series where finite mass corrections are neglected 
are given by
\begin{align}
m_b/\overline{m}_b
&=1+0.424413 \alpha_s^{(4)}(\overline{m}_b^2)+0.94005 (\alpha_s^{(4)}(\overline{m}_b^2))^2
+3.0385 (\alpha_s^{(4)}(\overline{m}_b^2))^3+12.647 (\alpha_s^{(4)}(\overline{m}_b^2))^4+\cdots , \non
m_c/\overline{m}_c
&=1+0.424413 \alpha_s^{(3)}(\overline{m}_c^2)+1.0456 (\alpha_s^{(3)}(\overline{m}_c^2))^2
+3.7509 (\alpha_s^{(3)}(\overline{m}_c^2))^3+17.438 (\alpha_s^{(4)}(\overline{m}_c^2))^4+\cdots . \label{naive}
\end{align}
See refs.~\cite{Ball:1995ni, Ayala:2014yxa, Hayashi:2021vdq} for discussions
why finite mass corrections and use of the 3-flavor coupling are 
crucial.
We can see that the two series of eq.~\eqref{masscorr}, where mass corrections are fully considered,
are close to the second series of eq.~\eqref{naive}.
Based on this observation, we use the $\mathcal{O}(\alpha_s^4)$ coefficient of 17.438
in our analysis, although finite mass corrections at $\mathcal{O}(\alpha_s^4)$
are not known yet.

\section{Result}
\label{sec:4}

Following the procedure explained above, we obtain
$m_{b, {\rm RF}}$ and $m_{c, {\rm RF}}$ using the four-loop order perturbative series.
We use the inputs $\overline{m}_b=4.18^{+0.03}_{-0.02}$~GeV, $\overline{m}_c=1.27 \pm 0.02$~GeV
and $\LMS=0.332 \pm 0.015$~GeV.
To determine $\bar{\Lambda}_{\rm RF}$ and $(\mu_{\pi}^2)_{\rm RF}$,
we also use experimental mass values $\overline{M}_{B, {\rm exp}}=5.313$~GeV and $\overline{M}_{D, {\rm exp}}=1.971$~GeV. 
The errors of the experimental values can be neglected.
We obtain
\begin{align}
\overline{\Lambda}_{\rm RF}&=0.495(15)_{\mu}(49)_{\overline{m}_b}(12)_{\overline{m}_c} (13)_{\alpha_s} (0)_{\rm f.m.}~{\rm GeV} , \non
(\mu_{\pi}^2)_{\rm RF}&=
-0.12(13)_{\mu}(45)_{\overline{m}_b}(11)_{\overline{m}_c} (4)_{\alpha_s} (0)_{\rm f.m.}~{\rm GeV}^2 .
\end{align}
The first error indicates higher order uncertainty. We estimated it by examining how
large the result is modified when we use $\tilde{r}(\tau)$ calculated with a different renormalization scale as $\tau^2 \to s \tau^2$.
We take $s=1/2$ or $2$. 
We can see that the higher order uncertainty for $\bar{\Lambda}$ is well below 
100~\%, which is expected when the $u=1/2$ renormalon is not subtracted. 
The uncertainty for $\mu_{\pi}^2$ is still large, but we infer based on our analysis
using an estimated $\mathcal{O}(\alpha_s^5)$ coefficient
that this does not indicate that renormalon is not subtracted but this is
because the order of the available series is not high enough.
The second, third and fourth errors come from the uncertainties of the input parameters.
The last error is regarding the finite mass corrections. We estimated it by
replacing the $\mathcal{O}(\alpha_s^3)$ coefficients in eq.~\eqref{masscorr} with that of 
the second series in eq.~\eqref{naive}, where the mass corrections are omitted.

We obtain our final results by combining all the errors in quadrature:
\begin{align}
\overline{\Lambda}_{\rm RF}&=0.495\pm0.053~{\rm GeV} , \non
(\mu_{\pi}^2)_{\rm RF}&=
-0.12\pm 0.23~{\rm GeV}^2 .
\end{align}
We quote other determinations for comparison.
Ref.~{\cite{FermilabLattice:2018est} gave
\begin{align}
\overline{\Lambda}_{\rm RF}&=0.435(31)~{\rm GeV} , \non
(\mu_{\pi}^2)_{\rm RF}&=0.05(22)~{\rm GeV}^2   ,
\end{align}
and ref.~ \cite{Ayala:2019hkn} gave
\be
\overline{\Lambda}_{\rm RF}=477(\mu)^{-8}_{+17}(Z_m)^{+11}_{-12}(\alpha_s)^{-8}_{+9}(\mathcal{O}(1/m_h))^{+46}_{-46}~{\rm GeV}   .
\ee 

\section{Conclusions}
\label{sec:5}

As high-order perturbative series are available, it is becoming necessary
to include nonperturbative effects in QCD calculations using the OPE.
To this end, the renormalon problem needs to be resolved.
We recently proposed a renormalon subtraction method using Fourier transform.
Here we applied our method to a determination of the HQET nonperturbative
matrix elements, $\bar{\Lambda}$ and $\mu_{\pi}^2$. 
Our determination is carried out by subtracting the $u=1/2$ renormalon
and, for the first time, the $u=1$ renormalon from the relation
between the pole and the ${\overline{{\rm MS}}}$ masses.
We finally mention that our method is applicable to general single-scale high energy QCD observables.

\section*{Acknowledgement}

This work is based on refs.~\cite{Hayashi:2020ylq,Hayashi:2021vdq},
which is done in collaboration with Yuuki Hayashi and Yukinari Sumino.
The author is grateful to them.
This work is supported by JSPS KAKENHI Grant Numbers JP19K14711 
and MEXT KAKENHI Grant Number JP18H05542.

\bibliographystyle{utphys}
\bibliography{reference}

\end{document}